# Excitation function of $^{27}$Al+ $^{20}$Ne reaction upto 166 MeV

Dibyasree Choudhury[a], Susanta Lahiri[a,b*],
[a]Saha Institute of Nuclear Physics, 1/AF Bidhannagar, Kolkata-700064, India
[b]Homi Bhabha National Institute, India

*Corresponding author: susanta.lahiri@saha.ac.in

**Abstract:**

Aluminium is used as energy degrader or catcher foil or as backing material in nuclear physics experiments. Excitations functions of $^{27}$Al+ $^{20}$Ne reaction in the projectile energy range 49-166 MeV were measured using stacked foil technique. Various evaporation residues such as $^{43,44}$Sc, $^{41}$Ar, $^{34m,38}$Cl, etc., were identified in the matrix.

1. Introduction

Use of aluminium as degrader foil is commonly encountered in nuclear physics experiments due to its easy availability, malleability, high stopping power, etc. Al is also frequently used as catcher foil or backing material. While an experiment is designed to study the reaction products with a particular target-projectile combination, the reaction products on the interaction of projectile with Al as degrader, catcher or backing might interfere with the experiment under study. Therefore it is of importance to know the reaction products arising from the interference of Al and various projectiles in wide energy range.

Recently Variable Energy Cyclotron Centre (VECC), Kolkata is equipped with $^{20}$Ne beam. In this paper, Al foils were irradiated with the heavy ion beam $^{20}$Ne from 53.5-170 MeV incident energy. Earlier research has been carried out using $^{27}$Al+$^{20}$Ne target-projectile combination. In 1984, Kox et al [1] studied the total cross section of $^{27}$Al+$^{20}$Ne reaction at 30 MeV/nucleon and reported a total reaction cross section of 2130±120 mb. Recently, the total charge-changing cross sections and partial cross sections for fragment production of 388 A MeV [2] and 400 A MeV [3] $^{20}$Ne ions in collision with H, C, Al and CH$_2$ targets using CR-39 detector were also studied. However, none of the works reported the production cross section of the radioisotopes produced from interaction of $^{20}$Ne on Al target. In this paper, the production of several radioisotopes such as $^{43,44}$Sc, $^{41}$Ar, $^{34m,38}$Cl, etc. was observed. Excitation function of $^{27}$Al+ $^{20}$Ne reaction was studied as a function of $^{20}$Ne energy. This experiment also marked the possibility of alternative production routes for above mentioned radioisotopes, some being of clinical significance such as $^{43,44}$Sc. Although the yield of radioisotopes produced in this experiment was not high enough for clinical applications, nevertheless, activity might be enough for tracer study in biological experiments.

2. Experimental

*2.1 Irradiation and data acquisition:* Total 8 targets were irradiated in respective stacks with 53.3-170 MeV incident $^{20}$Ne beam at VECC Kolkata, India. The $^{20}$Ne beam was produced from the ECR ion source. The exit energy of the projectile was calculated by the software SRIM [4]. The irradiation details are mentioned in Table 1. The incident beam was well-collimated on the target holder and the total charge of the incident particle was measured with the help of Faraday cup placed at the rear end of the target in conjunction with a current integrator (Danfysik). A series of off-line γ- spectra were taken for 7 days in a p-type HPGe detector (30%) of resolution 2.06 keV at 1.33 MeV in combination with a digital spectrum analyzer (DSA 1000, CANBERRA) and Genie 2K software (CANBERRA). The radionuclides were identified from their corresponding photo peak and decay data. The energy and efficiency calibration of the detector were performed using $^{152}$Eu ($T_{1/2}$ = 13.53 a), $^{133}$Ba ($T_{½}$ = 10.51 a), $^{60}$Co ($T_{½}$ = 5.2 a) and $^{137}$Cs ($T_{½}$ = 30.07 a) sources of known activity.

*2.2 Calculation of cross section:* The production cross sections of the radioisotopes were measured as a function of $^{20}$Ne energy and it was calculated from the activity at the end of bombardment (EOB), by using the following activation equation:

$$A = n\sigma(E)I(1 - e^{-\lambda T})$$

Where,  
    A= Activity (Bq) of a particular radionuclide at EOB  
    σ (E)= cross section of production of the radionuclide at incident energy E  
    I= Intensity of neon beam in particles/s  
    n= no.of atoms/cm$^2$  
    λ= disintegration constant  
    T= Duration of irradiation

*2.3 Error Calculation:* Uncertainty due to efficiency calibration was found to be insignificant (~0.2-0.6%) and hardly has any influence on the total uncertainty. Errors due to counting statistics were different for different isotopes: $^{43}$Sc (0.6-12%), $^{44m}$Sc (1.01-20.8%), $^{24}$Na (0.93-4.2%), $^{38}$Cl (5.11-12.6%), $^{34m}$Cl (0.38-13.26%), $^{42}$K (3.1-14.12%) and $^{41}$Ar (8.25-18.04%). Error in determining the target thickness was ~5%. The combined uncertainties due to beam current, incident beam energy, etc. were ~10%. The total error related to the cross-sectional measurement was 11.5 to 14.3%.

3. **Results and discussions**

Analysis of the γ- spectrum revealed the production of $^{43,44,44m}$Sc, $^{34m,38}$Cl, $^{24}$Na, $^{41}$Ar, $^{42}$K and $^{43}$K radioisotopes (Figure 1) in the matrix. Table-2 lists the produced radioisotopes, their decay properties, associated gamma-rays, and possible production mechanism. The production cross section of the radionuclides has been provided in Table 3 and plotted in Figure 2.

The $^{44}$Sc radioisotope may be directly produced from $^{27}$Al via ($^{20}$Ne,2pn) reaction or it may be produced from $^{44m}$Sc as a daughter product. The metastable state has a much longer half life and it feeds the ground state (IT 98.61%) and also decays to $^{44}$Ca (EC 1.39%). Therefore, we could measure the yield at EOB and production cross section only for $^{44m}$Sc and not for $^{44}$Sc.

## Conclusion

Attempt has been made to study the possible radioisotopes produced from Al target which is most commonly used as degrader foils. Several radioisotopes such as $^{43,44}$Sc, $^{24}$Na, $^{34m}$Cl, etc., were identified. Excitation function of $^{27}$Al+$^{20}$Ne reaction from 49-166 MeV has been reported. The calculation of theoretical cross section values by different reaction model codes such as EMPIRE3.2.2, PACE4, etc. is under progress.

## Acknowledgement

Authors are thankful to VECC cyclotron staffs for their cooperation during neon irradiation. We gratefully acknowledge the support from the research grants from SINP-DAE 12 Five year plan Trace, Ultratrace Analysis and Isotope Production (TULIP), Government of India.

**Table 1: Irradiation details**

| Target | Target thickness (mg/cm$^2$) | Incident energy (MeV) | Exit energy (MeV) | Energy at centre of mass (MeV) | Irradiation time (h) | Integrated Charge (μC) |
|---|---|---|---|---|---|---|
| Al1 | 2.0 | 170 | 162.4 | 166 | 3.95 | 846 |
| Al2 | 7.6 | 160 | 130 | 145 | 11.60 | 964 |
| Al3 | 7.6 | 145.5 | 113.4 | 129 | 14.75 | 589 |
| Al4 | 2.0 | 119.8 | 110.2 | 115 | 3.75 | 846 |
| Al5 | 1.5 | 102.6 | 94.7 | 99 | 3.75 | 846 |
| Al6 | 1.5 | 84.5 | 75.8 | 80 | 11.60 | 964 |
| Al7 | 3.5 | 75.8 | 54.2 | 65 | 14.75 | 589 |
| Al8 | 1.5 | 54.2 | 43.4 | 49 | 14.75 | 589 |

**Table 2: Nuclear characteristics of the radioisotopes produced
(https://www.nndc.bnl.gov/chart/chartNuc.jsp)**

| Radioisotope | Half-life | Decay mode (%) | Principal Gamma energies keV (I,%) | $E_{Threshold}$ (MeV) |
|---|---|---|---|---|
| $^{43}$Sc | 3.89 h | ε+β$^+$ (100) | 372.7 (23) | 32.6 |
| $^{44}$Sc | 3.92 h | ε+β$^+$ (100) | 1157.03 (99.9) | |
| $^{44m}$Sc | 58.6 h | IT (98.8) | 271.1 (86.7) | 15.7 |
| | | ε+β$^+$ (1.2) | 1157.03 (1.2) | |
| $^{42}$K | 12.3 h | β$^-$ (100) | 1524.7 (18) | 46.0 |
| $^{41}$Ar | 109.3 m | β$^+$ (100) | 1293.5 (99.1) | 62.1 |
| $^{34m}$Cl | 32.0 m | IT (44.6) | 146.3 (40.5) | |
| | | ε+β$^+$ (55.4) | 1176.6 (14.1) | - |
| $^{38}$Cl | 37.24 m | β$^-$ (100) | 1642.7 (32) | |
| $^{24}$Na | 14.95 h | β$^-$ (100) | 1368.6 (100) | |

**Table 3**
**Cross section of the produced isotopes at EOB at experimental conditions**

| Energy (at the centre of mass of the target) (MeV) | $^{43}$Sc | $^{44m}$Sc | $^{24}$Na | $^{34m}$Cl | $^{38}$Cl | $^{42}$K | $^{41}$Ar |
|---|---|---|---|---|---|---|---|
| | | | | mb | | | |
| 166 | 0.06±0.007 | 0.15±0.02 | 0.97±0.11 | 0±0 | 0±0 | 0±0 | 0±0 |
| 145 | 0.07±0.008 | 0.08±0.01 | 0.73±0.08 | 1.49±0.19 | 0±0 | 0.08±0.01 | 0.01±0.002 |
| 129 | 0.69±0.08 | 0.13±0.02 | 1.24±0.14 | 2.86±0.37 | 0.43±0.06 | 0.33±0.05 | 0.02±0.003 |
| 115 | 1.71±0.21 | 0.35±0.05 | 0.68±0.07 | 4.1±0.53 | 0±0 | 0.67±0.09 | 0.01±0.002 |
| 99 | 6.62±0.81 | 1.57±0.24 | 0.36±0.04 | 3.87±0.50 | 0±0 | 0.51±0.07 | 0±0 |
| 80 | 9.13±1.12 | 4.2±0.65 | 0.22±0.02 | 2.13±0.27 | 0±0 | 0.2±0.028 | 0±0 |
| 65 | 7.92±0.97 | 91.8±14.3 | 0±0 | 0±0 | 0±0 | 0±0 | 0±0 |
| 49 | 0.05±0.006 | 19.2±2.9 | 0±0 | 0±0 | 0±0 | 0±0 | 0±0 |

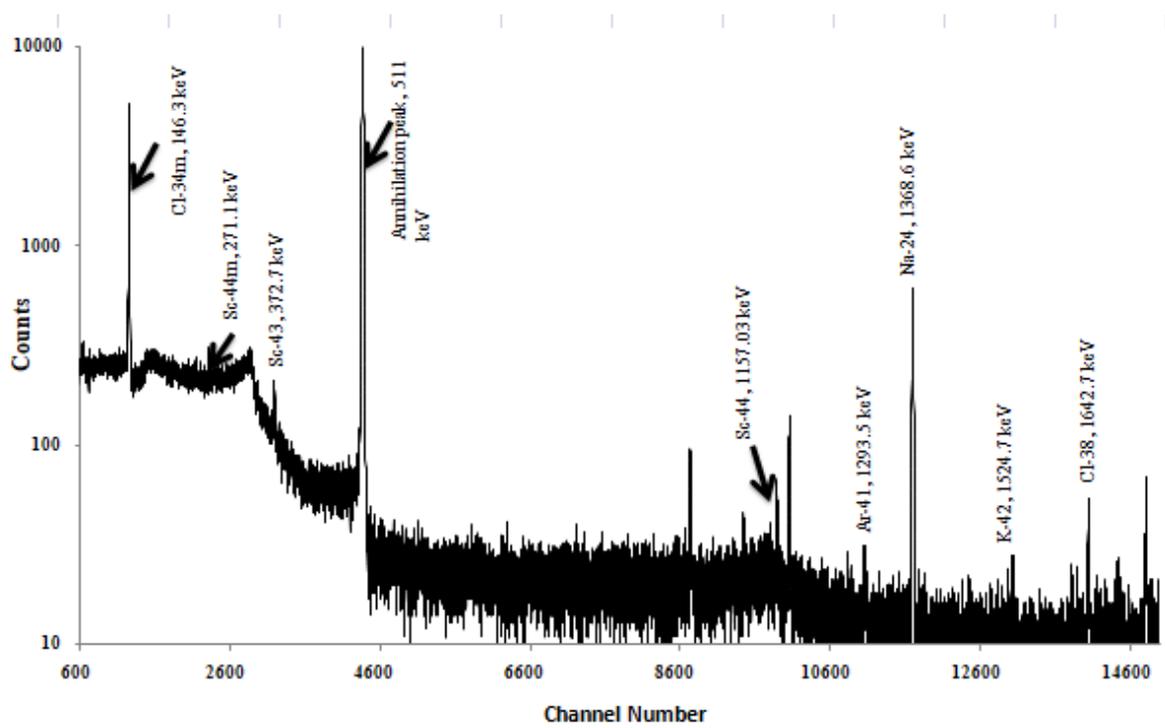

Fig.1: Gamma spectrum of Al foil irradiated at 160 MeV taken 1.0 h after EOB

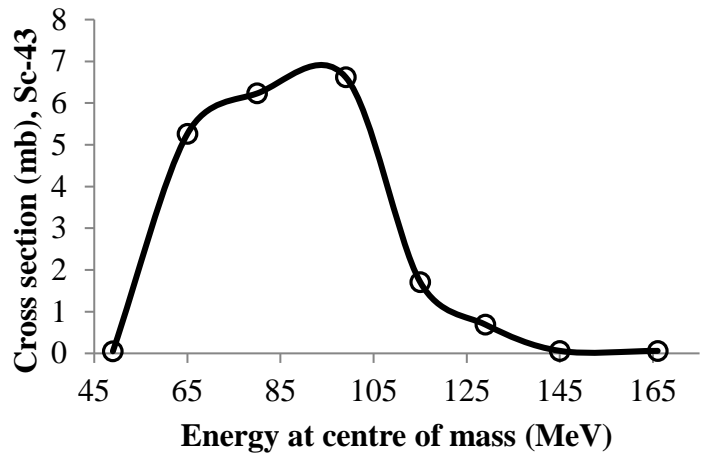
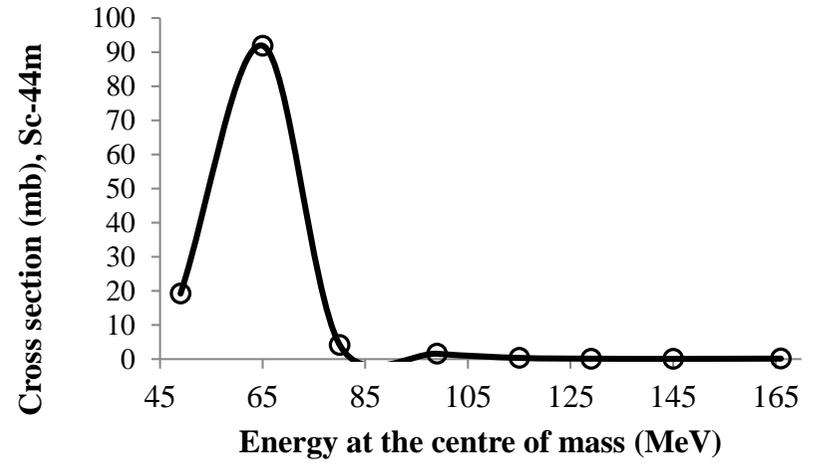
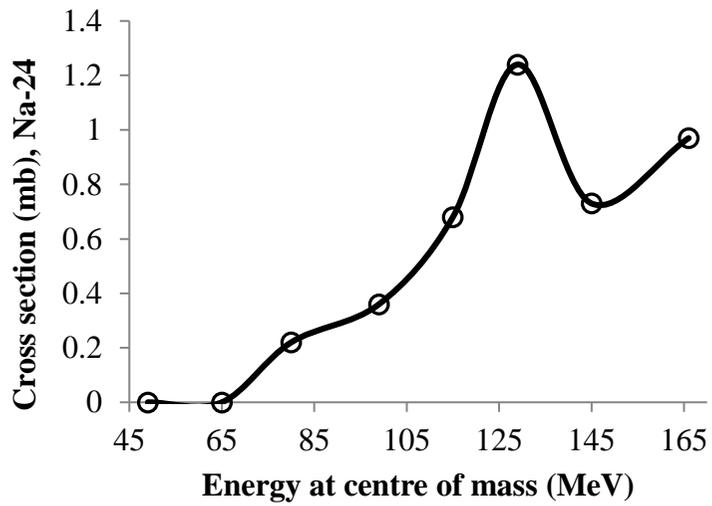
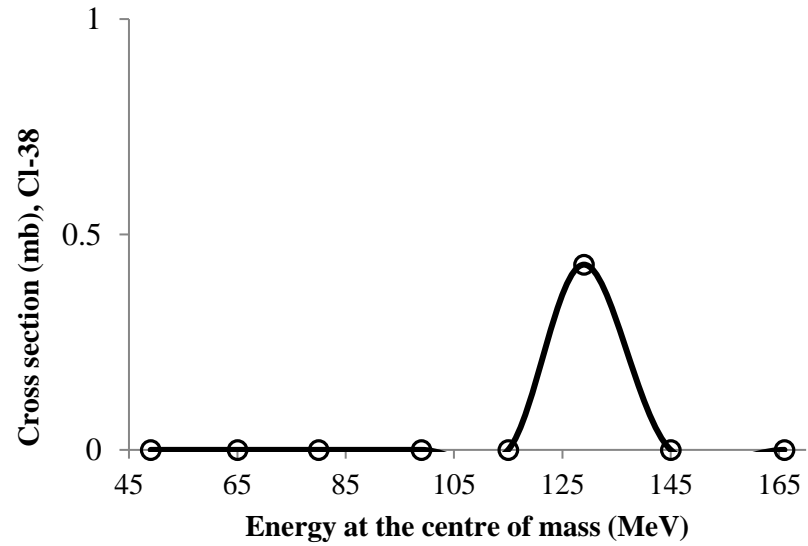

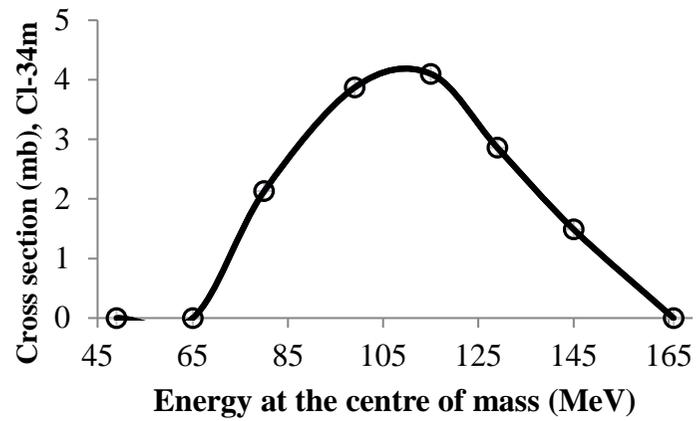
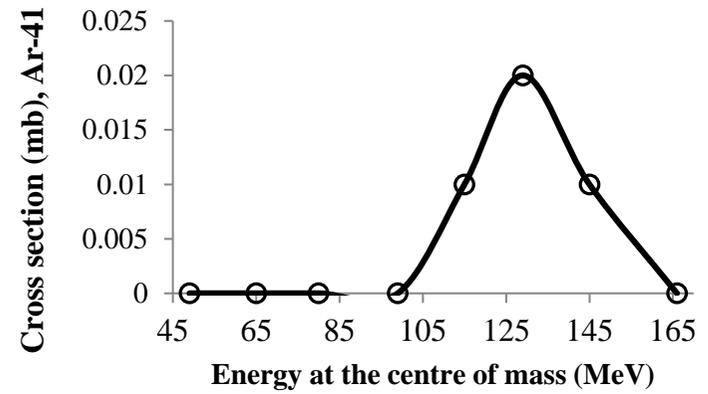
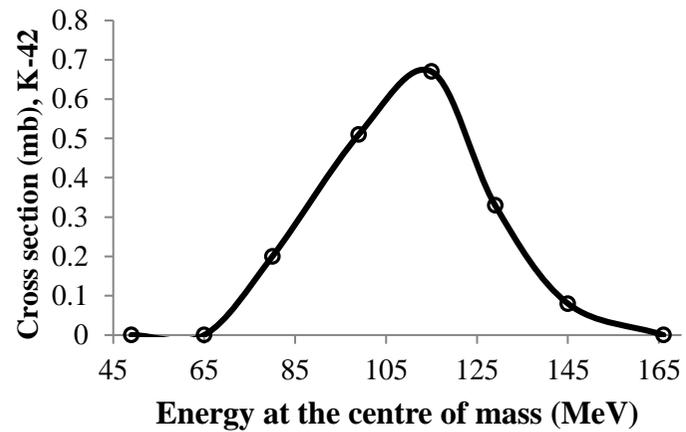

Fig 2: Cross section of $^{27}$Al+$^{20}$Ne reaction